\documentclass[aps,preprint,showpacs]{revtex4}

\usepackage[pdftex]{graphicx}
\usepackage{amssymb}
\usepackage{bm}
\begin{document}
\setcounter{page}{1}
\title[]{Non-Gaussianity in Nonminimally Coupled Scalar Field Theory}
\author{Seoktae \surname{Koh}}
\email{steinkoh@ewha.ac.kr}
%\thanks{Fax: +82-2-554-1643}
%\author{Young Sil \surname{Chang}}
\affiliation{Department of Science Education, Ewha Womans University, Seoul
120-750, Korea, and\\
Asia Pacific Center for Theoretical Physics, Pohang 790-894, Korea}
%\author{Jang \surname{Lee}}
%\affiliation{Department of Chemisty, HanKook University, Seoul
%135-703}
%\date[]{Received January 5 2004}

\begin{abstract}
We consider the non-Gaussianity of the nonlinear density perturbations in 
a single-field inflationary model when a scalar field couples nonminimally
with gravity. Gravity theories with a nonminimal coupling can be 
transformed into the Einstein gravity with canonical kinetic 
terms by using a suitable conformal transformation. We find that a
nonlinear generalization of the gauge invariant quantity $\zeta_i$  is 
invariant under the conformal transformation. With the help of this conformal
invariant property, we calculate the non-Gaussianity, 
which is characterized by a nonlinear
parameter $f_{NL}$, in nonminimal coupled scalar field theory. 
\end{abstract}



\maketitle

\section{INTRODUCTION}
The inflation scenario can explain successfully
 a seed for density perturbations which are
adiabatic and Gaussian, as well as the problems of 
Friedmann-Robertson-Walker(FRW) 
cosmology, such as the flatness and the horizon problems 
\cite{linde90, liddle00}. 
Also observational data fit the theoretical predictions within
linear perturbation theory to a good accuracy 
\cite{bardeen80,mukhanov92,liddle93}.
However, recent observations  with  high precision
({\it e.g.}, WMAP, PLANCK) \cite{komatsu03}
require more a accurate perturbation theory,  
beyond linear order, during the inflation period.
For example, if non-linear perturbations are taken 
into account, a non-Gaussian signal
is expected to be detected on the temperature anisotropy in future
expreiments \cite{acquaviva03, maldacena03}. 
Detection of the non-Gaussianity can give information
about the generation mechanism of density perturbations (such as
inflaton-, curvaton-, or inhomogeneous reheating scenario)
\cite{bartolo04}
and discriminate inflation models (such as single-field
\cite{acquaviva03,maldacena03,seery05} or multi-field
inflation \cite{seery05-2}). 

Generalized gravity theories beyond the Einstein gravity 
naturally
emerge from the fundamental physics theory, such as superstring/M-theory,
and are thought
to explain the present accelerating universe.
Non-Gaussianity in generalized gravity theory can also be calculated
using Hamilton-Jacobi formalism in Ref. \cite{koh05}.
In this paper, we calculate the non-Gaussianity during inflation 
when the scalar field is coupled to the gravitational curvature
nonminimally. Instead of the Hamilton-Jacobi approach in generalized
gravity theory,  which was adopted 
in Ref. \cite{koh05}, we use 
the conformal transformation method that transforms
nonminimally coupled scalar field theory into the canonical Einstein gravity. 
Further, we find that the nonlinear curvature perturbation quantity
on uniform field hypersurfaces (comoving hypersurfaces) 
is invariant under the conformal 
transformation.

\section{Curvature Perturbations in Nonminimally Coupled Scalar field Theory}
We start with the action for which the scalar field is coupled to the
Ricci scalar nonminimally:
\begin{eqnarray}
S = \int d^4 x \sqrt{-g} \left[
(1-\kappa\xi \phi^2)\frac{R}{2\kappa}-\frac{1}{2}g^{\mu\nu}\partial_{\mu}
\phi\partial_{\nu}\phi-V(\phi)\right],
\end{eqnarray}
where $\xi$ is a nonminimal coupling constant and $\kappa=8\pi G$.
We consider the Arnowitt-Deser-Misner (ADM) metric
\begin{eqnarray}
ds^2 = -N^2 dt^2 +\gamma_{ij}dx^i dx^j,
\end{eqnarray}
where $N$ and $\gamma_{ij}$ are a lapse function and a 3-spatial metric,
respectively.
The 3-spatial metric $\gamma_{ij}$ can be written as
\begin{eqnarray}
\gamma_{ij}=a^2(t)e^{2\alpha(t,{\bf x})}\delta_{ij}.
\end{eqnarray}
In this paper, we neglect the gravitational wave contribution
\cite{koh05, salopek90}.
The local Hubble parameter $H$ takes the form
\begin{eqnarray}
H \equiv \frac{\dot{a}}{Na}+\frac{\dot{\alpha}}{N}.
\end{eqnarray}
The gauge invariant quantity in nonlinear theory
was introduced in the literature \cite{salopek90,koh05, langlois05} 
and is given by
\begin{eqnarray}
\zeta_i = \partial_i \alpha - \frac{NH}{\dot{\phi}}\partial_i \phi.
\end{eqnarray}
If we expand $\zeta_i$ up to second order as $g=g_1+\frac{1}{2}
g_2$, then the linear and the second order gauge invariant quantities are
obtained in single field inflation model \cite{acquaviva03, malik04} as
\begin{eqnarray}
\zeta_1 &=&\alpha_1 -\frac{H_0}{\dot{\phi}_0}\delta \phi, \\
\zeta_2 &=& \alpha_2-\frac{H_0}{\dot{\phi}_0}\phi_2
+2\frac{H_0}{\dot{\phi}_0^2}\phi_1 \dot{\phi}_1
-2\frac{\phi_1}{\dot{\phi}_0}\dot{\alpha}_1 \nonumber \\
& & +\frac{H_0\phi_1^2}{\dot{\phi}_0^2}\left(\frac{\dot{H}_0}{H_0}
-\frac{\ddot{\phi}_0}{\dot{\phi}_0}\right),
\end{eqnarray}
where $\zeta_i =\partial_i \zeta$. Similarly to linear perturbation
theory, $\zeta_i$ can be understood as a curvature perturbation
on comoving hypersurfaces (or uniform field hypersurfaces) and 
is conserved when the perturbation length scales are larger than
the horizon scale for an adiabatic perturbation.

Nonlinear perturbations may generate a non-Gaussian signal
on the temperature anisotropy. To show this non-Gaussianity,
$\zeta$ can be decomposed into linear and nonlinear parts with a
nonlinear parameter $f_{NL}$\cite{komatsu01}:
\begin{eqnarray}
\zeta = \zeta_L +\frac{3}{5}f_{NL}(\zeta_L^2-\langle \zeta_L^2\rangle),
\label{fnl}
\end{eqnarray}
where $\zeta_L$ is a linear Gaussian perturbation that satisfies 
$\langle \zeta_L \rangle =0$.
With this definition, the 
non-vanishing component of the $\zeta({\bf x})$ bispectrum is 
\begin{eqnarray}
\langle \zeta_L({\bf k}_1)\zeta_L({\bf k}_2)\zeta_{NL}({\bf k}_3)\rangle
&=& 2 (2\pi)^3 \delta^{(3)}({\bf k}_1+{\bf k}_2
+{\bf k}_3) \nonumber \\
& &\times f_{NL}P_{\zeta}({\bf k}_1)P_{\zeta}({\bf k}_2),
\label{bispectrum}
\end{eqnarray}
where
\begin{eqnarray}
& &\langle \zeta_L({\bf k}_1)\zeta_L({\bf k}_2)\rangle
=(2\pi)^3P_{\zeta}({\bf k}_1)\delta^{(3)}({\bf k}_1+{\bf k}_2), \\
& &\zeta_{NL}({\bf k}) =
f_{NL}\biggl[\int \frac{d^3k^{\prime}}{(2\pi)^3}
\zeta_L({\bf k}+{\bf k}^{\prime})\zeta^{\ast}_L({\bf k}^{\prime}) 
\nonumber \\
& &~~~~~~~~~~~~~~~~~~~
-(2\pi)^3\delta^{(3)}({\bf k})\langle \zeta^2_L({\bf x})\rangle
\biggr].
\end{eqnarray}
WMAP data give an observational constraint on $f_{NL}$ such that
$-58< f_{NL}<134$ \cite{komatsu03}.

\section{conformal transformation} \label{conformal}
It is more convenient to deal with the equations of motion
 in the Einstein gravity with a minimally coupled
scalar field by using a suitable conformal transformation
\begin{eqnarray}
\hat{g}_{\mu\nu} = \Omega^2(\phi) g_{\mu\nu}.
\end{eqnarray}
Applying the conformal transformation with $\Omega^2=1-\kappa \xi \phi^2$
and introducing a new scalar field $\hat{\phi}$ to make a canonical
Lagrangian form as
\begin{eqnarray}
\hat{\phi} = \int d\phi F(\phi),
\end{eqnarray}
we obtain the action in the Einstein frame:
\begin{eqnarray}
S = \int d^4 x \sqrt{-\hat{g}}\left[\frac{1}{2\kappa}\hat{R}-
\frac{1}{2}\hat{g}^{\mu\nu}\partial_{\mu}\hat{\phi}
\partial_{\nu} \hat{\phi} -\hat{V}(\hat{\phi})\right],
\end{eqnarray}
where
\begin{eqnarray}
F(\phi) =\frac{\sqrt{1-\kappa\xi(1-6\xi)\phi^2}}{\Omega^2}, \quad
\hat{V}(\hat{\phi}) = \frac{V(\phi)}{\Omega^4}.
\end{eqnarray}

The $\zeta$ in linear perturbation theory has been verified  to be 
invariant under a conformal transformation in inflation theory
with a nonminimal coupling \cite{makino91} and in generalized gravity
cases \cite{hwang97}. The conformal invariance of $\zeta$ implies that the
power spectrum of $\zeta$ is the same in both frames. 
It is important to check whether the nonlinear generalization of $\zeta$-
$\zeta_i$ is also invariant under the conformal transformation. To do so,
we decompose the conformal factor $\Omega$ 
into a homogeneous part, which depends only on the time, 
and inhomogeneous part as
\begin{eqnarray}
\Omega(t,{\bf x}) = \Omega_0(t)e^{\omega(t,{\bf x})}.
\end{eqnarray}
Under this decomposition, 
we can easily check that the quantities in each frame are related as
\begin{eqnarray}
d\hat{t} &=& \Omega_0 dt, \quad \hat{a} = \Omega_0 a, \nonumber \\
\hat{N}&=& e^{\omega}N, \quad \hat{\alpha} = \alpha +\omega, \nonumber \\
\hat{H} &=& \frac{1}{\Omega}\left(H+\frac{\dot{\Omega}}{N\Omega}\right).
\end{eqnarray}
With these relations,  we find 
\begin{eqnarray}
\hat{\zeta}_i &=& \partial_i \hat{\alpha} 
-\frac{\hat{N}\hat{H}}{\dot{\hat{\phi}}}\partial_i \hat{\phi}  \\
&=&\partial_i \alpha - \frac{NH}{\dot{\phi}}\partial_i \phi = \zeta_i.
\end{eqnarray}
The nonlinear gauge invariant quantity $\zeta_i$ in the original
frame exactly coincides with that in the transformed Einstein frame
to nonlinear order. Therefore, 
the power spectrum and the bispectrum of $\zeta_i$ also coincide 
in each frame:
\begin{eqnarray}
\hat{P}_{\zeta} = P_{\zeta}, \quad
\langle \hat{\zeta}_{k_1}\hat{\zeta}_{k_2}\hat{\zeta}_{k_3}\rangle
=\langle \zeta_{k_1}\zeta_{k_2}\zeta_{k_3} \rangle.
\end{eqnarray}
From the definition of $f_{NL}$ in Eq. (\ref{fnl}), 
conformal invariance of $\zeta_i$ gives $\hat{f}_{NL} = f_{NL}$.

\section{Three-point correlation functions and Non-Gaussianity}
While the nonvacuum initial states can modify a four-point correlation
compared to the Gaussian statistics \cite{lesgourgues97}, 
nonlinear perturbations give
a nonzero three-point correlation that vanishes in Gaussian statistics.
Since the bispectrum in the original Jordan frame
is the same as that in the Einstein frame, as shown in
Sect. \ref{conformal},  
we may borrow the results of Maldacena \cite{maldacena03}.
The third order action in terms of $\zeta$ has been constructed 
in Ref. \cite{maldacena03} to calculate the bispectrum in the Einstein frame 
to the $\epsilon^2$ order by the solving constraint equations and 
is given as
\begin{eqnarray}
S_3 &=&\int d^4 x \biggl[\left(\frac{4\pi}{m_{pl}}\right)^2
\frac{\dot{\hat{\phi}}^4}{\hat{H}^4}
(\hat{a}^3 \dot{\hat{\zeta}}^2 \hat{\zeta} +\hat{a}
(\partial\hat{\zeta})^2 \hat{\zeta}) \nonumber \\
& &~~~~~~~~-\frac{4\pi}{m_{pl}}\frac{\dot{\hat{\phi}}^2}{\hat{H}^2}
\dot{\hat{\zeta}}(\partial \hat{\zeta} \cdot
\partial \hat{\chi})\biggr] = \int d^4 \mathcal{\hat{L}}_I,
\label{thirdorder}
\end{eqnarray}
where $\partial^2 \hat{\chi} = 
\frac{\dot{\hat{\phi}}^2}{2\hat{H}^2}\dot{\hat{\zeta}}$.
The slow-roll parameters $\hat{\epsilon}$ and $\hat{\eta}$ are defined by
\begin{eqnarray}
\hat{\epsilon} &=& -\frac{\dot{\hat{H}}}{\hat{H}^2}=-\frac{4\pi}{m_{pl}^2}
\frac{\dot{\hat{\phi}}^2}{\hat{H}^2}=\frac{m_{pl}^2}{16\pi}
\left(\frac{\hat{V}_{,\phi}}{\hat{V}}\right)^2 \equiv \hat{\epsilon}_V, \\
\hat{\eta} &=&-\frac{\ddot{\hat{\phi}}}{\hat{H}\dot{\hat{\phi}}}
=\frac{m_{pl}^2}{8\pi}\frac{\hat{V}_{,\phi\phi}}{\hat{V}}
-\frac{m_{pl}^2}{16\pi}\left(\frac{\hat{V}_{,\phi}}{\hat{V}}\right)^2 \\
&\equiv& \hat{\eta}_V-\hat{\epsilon}_V.  \nonumber 
\end{eqnarray}
We have defined another useful slow-roll parameter $\hat{\delta}$ as
\cite{tsujikawa04}
\begin{eqnarray}
\hat{\delta} = 
\frac{\dot{\hat{Q}}}{2\hat{H}\hat{Q}}, 
\quad \hat{Q}=\left(\frac{\dot{\hat{\phi}}}{\hat{H}}\right)^2.
\end{eqnarray}
Note that $\hat{\chi}$ is the order of $\hat{\epsilon}$ from the definition.

%Then the three-point correlation function of $\hat{\zeta}$ in the
%interacting picture
%\begin{eqnarray}
%\langle \hat{\zeta}^3 \rangle = \langle \hat{U}^{\dag}
%\hat{\zeta}_I^3 \hat{U}\rangle,
%\end{eqnarray}
%where
%\begin{eqnarray}
%& &\hat{U} = T \exp(-i\int^t_{t_0}\hat{H}_I(t^{\prime})
%dt^{\prime}), \nonumber \\
%& &\hat{\zeta}_I (t,{\bf k}) 
%=e^{i\hat{H}_0(t-t_0)}\hat{\zeta}(t_0,{\bf k})
%e^{-i\hat{H}_0(t-t_0)}, \nonumber \\
%& &\hat{\zeta}(t,{\bf k}) = \hat{U}^{\dag}(t,t_0)\hat{\zeta}_I(t,{\bf k})
%\hat{U}(t,t_0).
%\end{eqnarray}

From the third order action, Eq. (\ref{thirdorder}), 
the bispectrum for $\hat{\zeta}$ becomes
in the interacting picture \cite{maldacena03}
\begin{eqnarray}
\langle \hat{\zeta}_{{\bf k}_1}\hat{\zeta}_{{\bf k}_2}
\hat{\zeta}_{{\bf k}_3}\rangle
=(2\pi)^3\delta^3(\sum{\bf k}_i)\frac{(8\pi)^2}{m_{pl}^4}
\frac{\hat{H}^4}{\hat{\epsilon}}
\frac{1}{\Pi_i(2k_i^3)}\mathcal{\hat{A}},
\end{eqnarray}
where
\begin{eqnarray}
\mathcal{\hat{A}} &=& 2\frac{\ddot{\hat{\phi}}}{\dot{\hat{\phi}}\hat{H}}
\sum_i k_i^3
+\frac{\dot{\hat{\phi}}^2}{\hat{H}^2}\biggl[\frac{1}{2}\sum_i k_i^3
+\frac{1}{2}\sum_{i\neq j}k_i k_j^2 \nonumber \\
& & +\frac{4}{k_{tot}}\sum_{i>j}k_i^2 k_j^2\biggr], \\
k_{tot} &=& k_1+k_2+k_3.
\end{eqnarray}
Then, we can read off the nonlinear parameter $\hat{f}_{NL}$ from the 
relations om Eq. (\ref{bispectrum}):
\begin{eqnarray}
\hat{f}_{NL} \simeq \frac{5}{6}(\hat{\eta}-2\hat{\epsilon}) 
= -\frac{5}{6}(\hat{\delta} +\hat{\epsilon}).
\end{eqnarray}
If we get back to the original Jordan frame, then
 from the fact that $\hat{f}_{NL}=f_{NL}$,
\begin{eqnarray}
\hat{f}_{NL} = -\frac{5}{6}(\hat{\delta}+\hat{\epsilon})=f_{NL}.
\end{eqnarray}
We can easily find the relations of the slow-roll parameters between
the Einstein frame  and the original Jordan frame:
\begin{eqnarray}
\hat{\epsilon} &=&\frac{\epsilon+\beta}{1+\beta}
-\frac{\dot{\beta}}{H(1+\beta)^2}, \\
\hat{\delta} &=&\frac{\delta - \beta}{1+\beta},
\end{eqnarray}
where
\begin{eqnarray}
\hat{Q} &=& \frac{Q}{\Omega^2}, \quad Q \equiv \frac{\dot{\phi}^2
+6\dot{\Omega}^2}{(H+\dot{\Omega}/\Omega)^2}, \nonumber \\
\beta &=& \frac{\dot{\Omega}}{H\Omega}.
\end{eqnarray}
In the slow-roll approximation, $\dot{\beta} \simeq 0$ \cite{tsujikawa04},
and for nonminimally coupled scalar field theory,
\begin{eqnarray}
1+\beta = 1-\frac{\kappa \xi \phi}{H\Omega^2}\dot{\phi}.
\end{eqnarray}
Then,
\begin{eqnarray}
f_{NL} = -\frac{5}{6}\frac{1}{1+\beta}(\delta +\epsilon).
\end{eqnarray}
The nonminimal coupling constant $\xi$ is restricted to a range of 
$|\xi|<10^{-3}$ \cite{koh05-2} 
to have the number of $e$-folds needed for 
 sufficient inflation \footnote{
Although for $\xi<0$, $|\xi|\gg 1$ is possible to give a sufficient
number of  $e$-folds, the initial value of the scalar field, $\phi_0$,
must be greater than a critical value $\phi_m \equiv m_{pl}/\sqrt{8\pi
|\xi|}$ \cite{koh05-2}}. 
With this constraint on $\xi$, 
the effect of nonminimal coupling, $1/(1+\beta)$, does not 
contribute much to $f_{NL}$ compared to the minimal coupling case.
Thus, as long as $\epsilon, \delta \ll 1$, non-Gaussian signal of the
single field slow-roll inflation model in a nonminimally coupled scalar
field theory is difficult to observe by using Cosmic Microwave 
Background (CMB) experiments as in the
Einstein gravity.

\section{CONCLUSIONS}
We have considered non-Gaussianity when the scalar field is 
nonminimally coupled to gravity. It is convenient to deal with
the equations of motion in a Einstein frame in which the scalar fields
are minimally coupled to the gravity.
The action in the nonminimally coupled scalar field theory is transformed 
into the Einstein gravity with canonical kinetic terms by using a
suitable conformal transformation. 

In linear perturbation theory, the gauge invariant quantity
$\zeta$, which is constant on uniform energy density hypersurfaces 
for superhorizon scales,
or $\mathcal{R}$ on comoving hypersurfaces is well known to be invariant 
under the conformal transformation \cite{hwang97, makino91}.
We have found that the gauge invariant quantity in nonlinear perturbation
theory $\zeta_i$ is also invariant under the conformal transformation.
Conformal invariance of $\zeta_i$ proves that the power spectrum and 
bispectrum of $\zeta_i$ are also conformally invariant. These 
facts reveal that the nonlinear parameters $f_{NL}$ in the two
frames should be the same.
Therefore, if we use Maldacena's result \cite{maldacena03}
in the Einstein frame, we can calculate 
the non-Gaussianity in nonminimally coupled scalar
field theory via a conformal transformation.

For $|\xi| <10^{-3}$, which is needed to ensure a sufficient
number of $e$-folds,
the effect of nonminimal coupling does not contribute much to the
nonlinear parameter $f_{NL}$ compared to the minimally coupling case.
As in the single field inflation model in Einstein gravity, a non-Gaussian
signal on the temperature anisotropy in nonminimally coupled scalar field
theory will be difficult to observe in future CMB experiments.

\begin{acknowledgments}
This work was supported by a Korea Research Foundation Grant
(KRF-2004-037-C00014).
\end{acknowledgments}

%\begin{references}

%\end{references}

\end{document}